\documentclass{emulateapj}
\slugcomment{Accepted: June 9, 2016}
\shorttitle{Injection to rapid DSA at perpendicular shocks in partially ionized plasmas}

\shortauthors{Yutaka Ohira}

\begin{document}

\title{INJECTION TO RAPID DIFFUSIVE SHOCK ACCELERATION AT PERPENDICULAR SHOCKS IN PARTIALLY IONIZED PLASMAS}

\author{Yutaka Ohira}

\begin{abstract}
We present a three-dimensional hybrid simulation of a collisionless perpendicular shock in a partially ionized plasma for the first time.
In this simulation, the shock velocity and the upstream ionization fraction are $v_{\rm sh} \approx 1333~{\rm km/s}$ and $f_{\rm i}\sim0.5$, that are typical values for isolated young supernova remnants in the interstellar medium. 
We confirm previous two-dimensional simulation results that downstream hydrogen atoms leak into the upstream region,  
they are accelerated by the pickup process in the upstream region, 
and large magnetic field fluctuations are generated both in the upstream and downstream regions. 
In addition, we find that the magnetic field fluctuations have three-dimensional structures and the leaking hydrogen atoms are injected to the diffusive shock acceleration at the perpendicular shock after the pickup process. 
The observed diffusive shock acceleration can be interpreted as the shock drift acceleration with scattering.
Particles are accelerated to $v\sim 100~v_{\rm sh}\sim0.3~c$ within $\sim 100$ gyroperiods in this simulation. 
The acceleration time scale is faster than that of the diffusive shock acceleration in parallel shocks. 
Our simulation results suggest that supernova remnants can accelerate cosmic rays to $10^{15.5}~{\rm eV}$ (the knee) during the Sedov phase. 
\end{abstract}


\affil{Department of Physics and Mathematics, Aoyama Gakuin University, 
5-10-1 Fuchinobe, Sagamihara 252-5258, Japan; ohira@phys.aoyama.ac.jp}
\section{INTRODUCTION}
The origin and acceleration mechanism of Galactic cosmic rays (GCRs) are long standing problems in the Astrophysics. 
Supernova remnants (SNRs) are believed to be the origin of GCRs, which is supported by recent gamma-ray observations \citep{ohiraetal11,ackermann13}. 
One of famous acceleration mechanisms is diffusive shock acceleration (DSA) \citep{axford77,krymsky77,bell78,blandford78}.  
The DSA theory assumes particles moving around a shock diffusively. 
Then, some of the diffusive particles move back and forth between the upstream and downstream regions many times. 
As a result, the diffusive particles are accelerated by the shock. 
The DSA theory predicts that the momentum spectrum of the accelerated particles has a power law form, that is almost consistent with the GCR spectrum once the escape from the sources and propagation effects are accounted for  \citep{ohiraetal10}.  
So that DSA appears to be the most plausible acceleration mechanism of GCRs. 

However, there are several problems for DSA. 
One is the injection problem. 
The DSA theory does not tell us how many particles are injected to DSA from thermal particles. 
The angle between the magnetic field and the shock normal, $\theta_{\rm Bn}$, is thought to be the most important parameter for the injection to DSA. 
The injection to DSA is expected to be more efficient for parallel shocks ($\theta_{\rm Bn}=0^{\circ}$) than for perpendicular shocks ($\theta_{\rm Bn}=90^{\circ}$) because the magnetic field perpendicular to the shock normal prevents charged particles from returning back to the upstream region \citep{ellison95}. 
\citet{caprioli14} showed by hybrid simulations that the injection efficiency markedly decreases with $\theta_{\rm Bn}$ for $\theta_{\rm Bn}\gtrsim 50^{\circ}$ (see Figure~3 of \citet{caprioli14}). 

Another problem is the knee problem. 
Since the CR spectrum observed at the Earth has a spectral break at the energy of $10^{15.5}~{\rm eV}$ (this is, so called, the knee), the energy of the knee is thought to be the maximum energy of CR protons accelerated by DSA at SNR shocks. 
However, if we assume a parallel shock with a typical magnetic field strength (a few $\mu$G) of the interstellar medium (ISM), the maximum energy of DSA in SNRs cannot reach the knee energy because the acceleration time scale of DSA is too large at the parallel shocks \citep{lagage83}. 
If magnetic fields are amplified to a few hundreds of ${\rm \mu}$G, 
the acceleration time scale becomes small, so that SNRs can accelerate CRs to the knee energy during the free expansion phase of SNRs. 
X-ray observations of young SNRs suggest the magnetic field amplification in the downstream regions \citep{vink03, berezhko03, bamba05, uchiyama07}. 
So far, some amplification mechanisms of magnetic fields have been proposed and investigated by simulations. 
However, what is the dominant mechanism as the magnetic field amplification remains an open question. 

Another proposed solution for the knee problem is DSA at the perpendicular shocks \citep{jokipii87}. 
Accelerated particles propagate a short distance from the shock surface for the perpendicular shock compared with the parallel shock because particles easily propagate to the shock normal direction for the parallel shock. 
As a result, the time scale of the back and forth motion around the shock front, that is, the acceleration time scale becomes small for the perpendicular shock. 
The fast acceleration at the perpendicular shock was confirmed by test-particle simulations \citep{decker86,takahara91,naito95,ellison95,giacalone05b,takamoto15}. 
However, as mentioned above, the injection to DSA is thought to be difficult for the perpendicular shock. 
In fact, \citet{caprioli14} showed by two- and three-dimensional hybrid simulations that only the acceleration via shock drift acceleration (SDA) last for a very short time in a pure perpendicular shock because particles are advected downstream, 
while some particles are injected to DSA and accelerated for a long time in a parallel shock. 
\citet{giacalone05a} showed by two-dimensional hybrid simulations 
that particles are injected to DSA (or multiple SDA) if there is magnetic turbulence in the upstream region of a globally perpendicular shock, but the upstream magnetic turbulence is not generated by the self-consistent manner and the acceleration time scale is not so small. 
Therefore, the upstream turbulence, the injection to DSA, and the fast acceleration of particles have never been simultaneously demonstrated by a self-consistent simulation.

Early studies for DSA at the perpendicular shock implicitly assumed that upstream plasmas are fully ionized. 
However, it is well known that the ISM and ejecta of SNRs are not always fully ionized. 
In fact, the observed line profile of ${\rm H_{\alpha}}$ emission from SNRs supports that some SNR shocks are propagating to the partially ionized ISM \citep{chevalier78,katsuda16}. 
It was proposed that many plasma instabilities are excited by the ionization around a collisionless shock in the partially ionized plasma \citep{raymond08,ohiraetal09,ohira10,ohira14}. 
\citet{ohira13} performed the first two-dimensional hybrid simulation of a collisionless shock wave propagating into a partially ionized plasma and showed that the ionization of hydrogen atoms excites plasma instabilities both in the upstream and downstream regions. 
In addition, the simulation showed that downstream hydrogen atoms leak into the upstream region. 
In the upstream region, density fluctuations are excited by the ionization of leaking neutral particles from the downstream region \citep{ohira13,ohira14}. 
Moreover, \citet{ohira16} performed a similar simulation to that of \citet{ohira13} but the Alfv{\'e}n Mach number and the shock velocity are higher and smaller, respectively. 
Then, very large fluctuations of magnetic fields and density are generated in the upstream and downstream regions.  
Although the leaking neutral particles are accelerated by the pickup process in the upstream region, no particles are injected to DSA in the two-dimensional hybrid simulations of \citet{ohira13,ohira16}. 
Since the particle diffusion perpendicular to the magnetic field line is suppressed in the two-dimensional system \citep{jokipii93,giacalone94}, 
a three-dimensional simulation is needed to understand whether the injection to DSA is possible or not in the perpendicular shock. 

In this paper, we perform the first three-dimensional hybrid simulation of collisionless shocks propagating to the partially ionized ISM. 
We then show that the upstream turbulence is self-consistently generated, leaking hydrogen atoms are injected to DSA, and the particles are rapidly accelerated by DSA at the perpendicular shock. 
In Section \ref{sec:2}, we first briefly summarize our hybrid simulation. 
Then, we show results of the three-dimensional hybrid simulation in Section \ref{sec:3}.
Finally, we discuss the acceleration time scale and spectral index of DSA at the perpendicular shock in Section \ref{sec:4} and summarize in Section~\ref{sec:5}.

\section{HYBRID SIMULATIONS}
\label{sec:2} 
Hybrid simulations compute the induction equation for the magnetic field with the generalized Ohm's law and the nonrelativistic equation of motion for many protons but electrons are treated as a massless fluid. 
In addition, our hybrid simulation solves the ionization and the free-streaming motion of hydrogen atoms \citep{ohira13,ohira16}. 
We consider the charge exchange and collisional ionization with protons, electrons, and hydrogen atoms as the ionization of hydrogen atoms. 
To compute the collisional ionization with electrons, we need to assume the velocity  distribution of electrons because the hybrid simulation does not solve the dynamics of electrons. 
In this simulation, we assume the Maxwell distribution with a temperature, $T_{\rm e}$ \citep[see][for details]{ohira16}. 
The ionization time scale of hydrogen atoms in typical isolated young SNRs is about $10^5$ times longer than the gyro period of protons. 
Since the huge gap of time scale is still challenging for current supercomputers, 
all cross sections are boosted by a factor of $2\times10^3$ in this paper.
Then, the ionization time scale in this simulation is about $50~\Omega_{\rm cp}^{-1}$, 
where $\Omega_{\rm cp}$ is the proton cyclotron frequency.  
Furthermore, in order to reduce the simulation box we put an upstream boundary at the sufficiently far upstream region, where leaking neutral hydrogen atoms are artificially ionized. 

Simulation particles are injected with a drift velocity parallel to the x direction at the left boundary, $x=0$, and specularly reflected at the right boundary $x=L_{\rm x}$. 
As time goes on, a perpendicular shock propagates to the left boundary and 
the mean velocity becomes zero in the downstream region. 
Accordingly, the simulation frame is the downstream rest frame. 
We impose the periodic boundary condition in both the $y$ and $z$ directions. 
The simulation box size is $L_{\rm x} \times L_{\rm y} \times L_{\rm z}= 8400~c / \omega_{\rm pp} \times 200~c / \omega_{\rm pp}\times 200~c / \omega_{\rm pp}$, where $c$ and $\omega_{\rm pp}$ 
are the speed of light and upstream plasma frequency of protons, respectively. 
The cell size and time step are set to 
$\Delta x = \Delta y = \Delta z =1~c / \omega_{\rm pp}$ and $\Delta t=10^{-3}~\Omega_{\rm cp}^{-1}$. 

To describe the upstream plasma, 8 protons and 8 hydrogen atoms per cell are used and the uniform magnetic field parallel to the $y$ direction is imposed, $\mbox{\boldmath $B$}_0=B_0 \mbox{\boldmath $e$}_{\rm y}$. 
Parameters of the upstream plasma are as follows: 
The ionization degree is $0.5$. 
The ratio of the particle pressure to the magnetic pressure is 
$\beta_{\rm p}=\beta_{\rm H}=0.5$ for protons and hydrogen atoms.  
The drift velocity parallel to the $x$ direction is $v_{\rm d}=20~v_{\rm A}=1000~{\rm km/s}$, 
where $v_{\rm A}=B_0/\sqrt{4\pi \rho_{\rm p,0}}$ is the Alfv{\'e}n velocity 
and $\rho_{\rm p,0}$ is the proton mass density in the far upstream region. 
For the electron temperature, we assume $T_{\rm e}=0$ and $0.01m_{\rm p}v_{\rm d}^2/3~(=0.01 \times (3/16)m_{\rm p} v_{\rm sh}^2)$ in the upstream and downstream region, respectively, where $m_{\rm p}$ is the proton mass. 
The factor of $0.01$ is based on several studies of electron heating in collisionless shocks \citep{ghavamian07,ohira07,ohira08,rakowski08,van08,laming14,vink15}.
We have not investigate the electron temperature dependence, that is future work. The number of leaking hydrogen atoms depends on the electron temperature 
\citep[e.g.][]{morlino12}. 

\section{SIMULATION RESULTS}
\label{sec:3}
\begin{figure}
\includegraphics[scale=0.7]{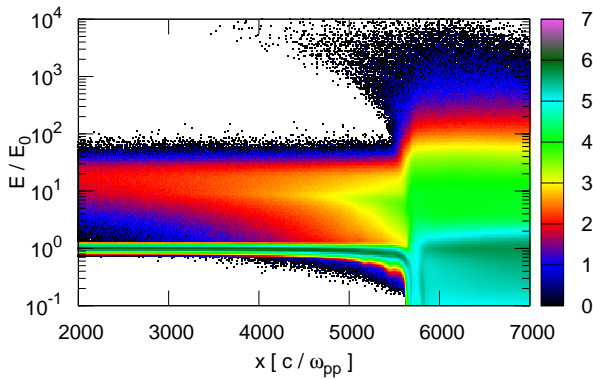}
\caption{Phase space of protons at $t=400~\Omega_{\rm cp}^{-1}$. The left regions ($x<5700~c/\omega_{\rm pp}$) are upstream regions. The vertical and horizontal axises show the kinetic energy normalized by the upstream kinetic energy, $E/E_0$ and the x coordinate, respectively. The color shows the phase space density in logarithmic scale. 
\label{fig:1}}
\end{figure}

\subsection{Phase space of protons}
We show the phase space of protons at $t=400~\Omega_{\rm cp}^{-1}$ in Figure~~\ref{fig:1}, 
where the vertical and horizontal axises show the kinetic energy, $E$, normalized by the upstream kinetic energy, $E_0=m_{\rm p}v_{\rm d}^2/2$, and the x coordinate, respectively. 
The shock front is propagating to the left direction and located at $x\approx 5700~c/\omega_{\rm pp}$.
The shock velocity is $v_{\rm sh} \approx 26.7~v_{\rm A}=37.7~v_{\rm A,tot}=1333~{\rm km/s}$ in the upstream rest frame, 
where $v_{\rm A,tot}=B_0/\sqrt{4\pi(\rho_{\rm p,0}+\rho_{\rm H,0})}$ is the Alfv{\'e}n velocity defined by the total mass density.  
The upstream cold component with $E\sim E_0$ and the upstream hot component with $E\sim10E_0$ are the upstream protons and pickup ions generated by the ionization of hydrogen atoms leaking from the downstream region. 
For shocks propagating to partially ionized plasmas, the hot hydrogen atoms that can leak into the upstream region are generated by the charge exchange process in the downstream region \citep{lim96,blasi12,ohira12,ohira13,ohira16}. 
In this simulation, the flux of leaking hydrogen atoms is a few of percent of the upstream hydrogen flux in the shock rest frame. 
In addition, there are protons accelerated to $E \sim 10^4E_0$ around the shock front, that were not observed in previous two-dimensional simulations \citep{ohira13,ohira16}.
Since two-dimensional simulations cannot accurately follow the perpendicular diffusion to the magnetic field line but three-dimensional simulations can do it \citep{jokipii93,giacalone94}, 
we can observe the particle acceleration around the shock front for the first time in this simulation. 
We discuss the acceleration mechanism in Section~\ref{sec:3.3}.

\subsection{Three dimensional structures}

\begin{figure*}
\centering
\includegraphics[width=0.7\textwidth]{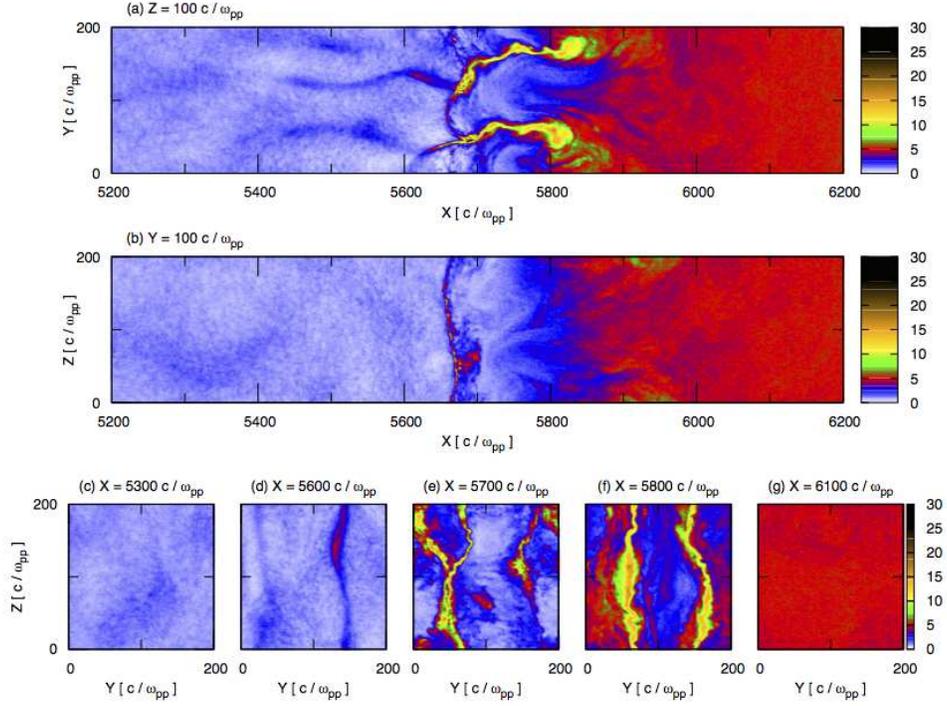}
\caption{Three-dimensional structures of the proton density normalized by the far upstream density, $\rho_{\rm p}/\rho_{\rm p,0}$, at $t=400~\Omega_{\rm cp}^{-1}$. 
The panels (a) and (b) show two-dimensional slices of the three-dimensional simulation box at $z=100~c/\omega_{\rm pp}$ and $y=100~c/\omega_{\rm pp}$, respectively. 
The panels (c)--(g) show slices through the y-z plane at $x=5300, 5600, 5700, 5800, {\rm and}~6100~c/\omega_{\rm pp}$, respectively. The direction of the upstream uniform magnetic field is the y direction. 
\label{fig:2}}
\end{figure*}
\begin{figure*}
\centering
\includegraphics[width=0.7\textwidth]{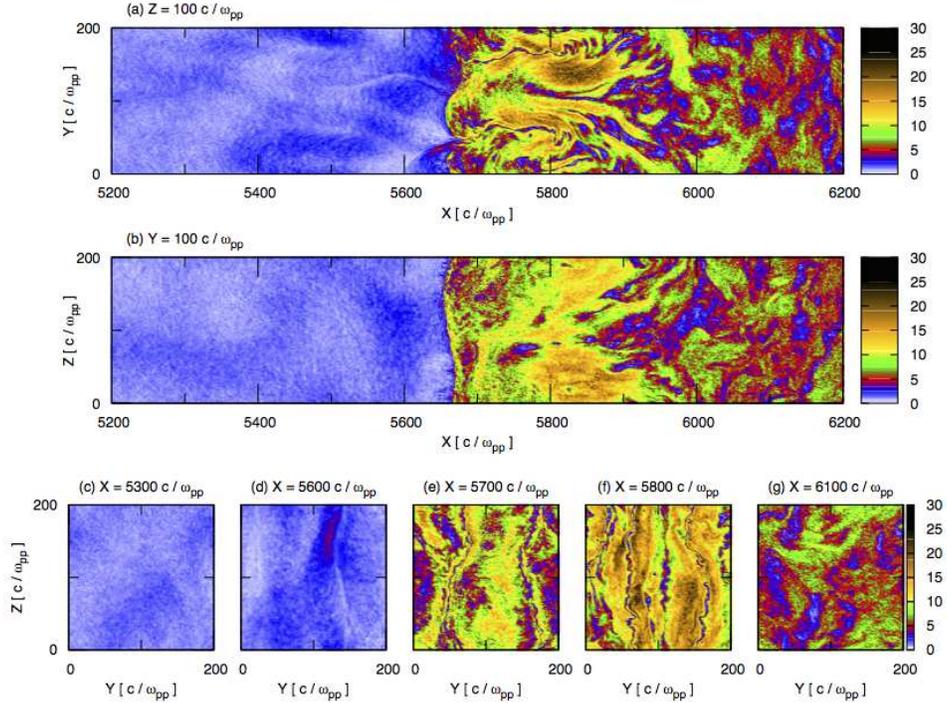}
\caption{The same as Figure~\ref{fig:2}, but for the magnetic field strength normalized by the upstream value, $|B|/B_0$. 
\label{fig:3}}
\end{figure*}

Figures~\ref{fig:2} and \ref{fig:3} show three dimensional structures at  $t=400~{\Omega_{\rm cp}}^{-1}$ for the proton density, $\rho_{\rm p}/\rho_{\rm p,0}$, and the magnetic field strength, $|B|/B_{0}$, respectively. 
In the Figures~\ref{fig:2} and \ref{fig:3}, the panels (a) and (b) show two-dimensional slices of the three-dimensional simulation box at $z=100~c/\omega_{\rm pp}$ and $y=100~c/\omega_{\rm pp}$, respectively. 
The panels (c)--(g) show slices through the y-z plane at $x=5300, 5600, 5700, 5800, {\rm and}~6100~c/\omega_{\rm pp}$, respectively. 
Shock structures averaged over the y and z directions is very similar to Figure 2 of \citet{ohira16}. 

In the upstream region ($x=5300~c/\omega_{\rm pp}$), variations are almost isotropic and the density is well correlated with the magnetic field strength. 
Around the shock front ($x=5600, 5700, {\rm and}~5800~c/\omega_{\rm pp}$), 
the proton density and magnetic field strength are very anisotropic and the density is anticorrelated with the magnetic field strength. 
In the far upstream region, the fast magnetosonic mode is excited by the leaking neutral particles \citep{ohira14}, so that there is correlation between the density and magnetic field strength. 
However, the fastest growing instability becomes the Alfv{\'e}n ion cyclotron instability around the shock front because there are many pickup ions around the shock front compared with the far upstream region \citep{ohira16}. 
The Alfv{\'e}n ion cyclotron instability is driven by the temperature anisotropy of the pickup ions and 
excites Alfv{\'e}n waves propagating along the magnetic field line. 
The Alfv{\'e}n waves with large amplitude push plasmas along the magnetic field line, so that the high density sheets are generated at regions where the amplitude of the Alfv{\'e}n waves is small. 
This is the reason why the anticorrelation appears. 
The strong turbulence is generated by interactions between the high density sheets and  the shock front. 
As a result, magnetic fields are strongly amplified to the equipartition level by the turbulence in the downstream region. 
In the far downstream region ($x=6100~c/\omega_{\rm pp}$), the high density sheets 
are disrupted by the downstream turbulence, so that variations become almost isotropic. 

These behaviors were observed in the previous two-dimensional simulations for the high Alfv{'e}n Mach number shock \citep{ohira16}. 
Therefore, we confirm that two-dimensional simulations can approximately follow actual behaviors of  the density and magnetic fields on the xy plane. 
However, Figures~\ref{fig:2} and \ref{fig:3} show that there are three-dimensional structures. 
This is crucial for the injection to DSA. 
In fact, we observe the particle acceleration up to $10^4~E_0$ in the three-dimensional simulation (see Figures~\ref{fig:1} and \ref{fig:4}), but do not observe it in the two-dimensional simulation. 
In the next subsection, we show the energy spectrum of protons and trajectories of accelerated particles for the three-dimensional simulation. 

\subsection{Energy spectrum and particle trajectory}
\label{sec:3.3}
\begin{figure}
\includegraphics[scale=0.7]{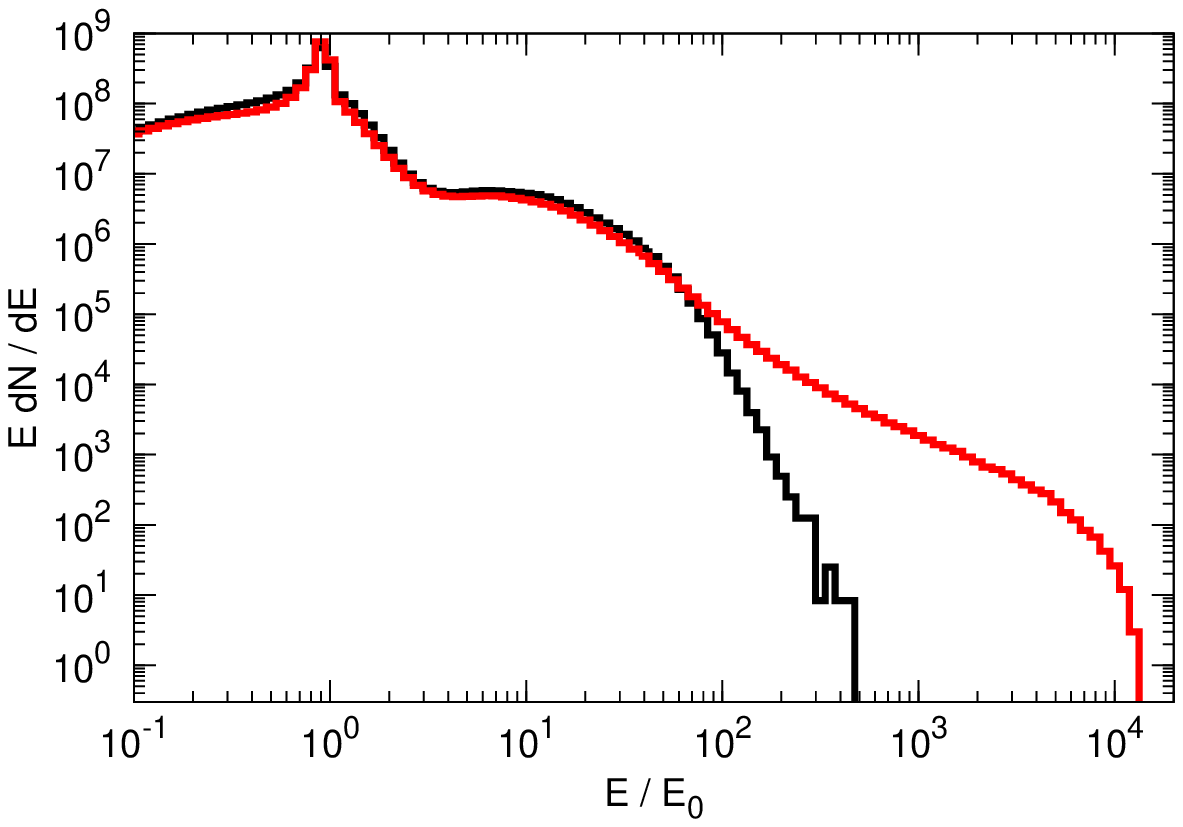}
\caption{Energy spectrum of protons in the region of $2000~c/\omega_{\rm pp}\le x\le 7000~c/\omega_{\rm pp}$ at $t=400~\Omega_{\rm cp}^{-1}$. 
The black and red histograms show the energy spectra for two- and three-dimensional simulations, respectively. \label{fig:4}}
\end{figure}
\begin{figure}
\includegraphics[scale=0.8]{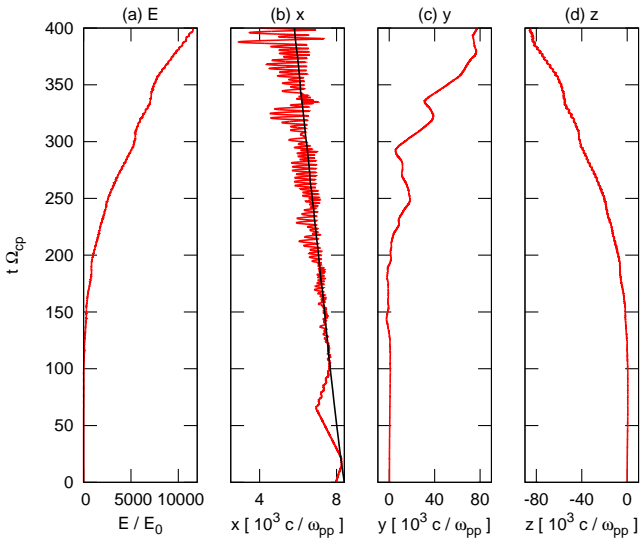}
\caption{Trajectory of an accelerated particle. The panel (a) shows the time evolution of energy, and the panels (b), (c), and (d) show particle trajectories in the x, y, and z coordinates, respectively. The back curve in the panel (b) shows the mean position of the shock front. 
\label{fig:5}}
\end{figure}
\begin{figure}
\includegraphics[scale=0.7]{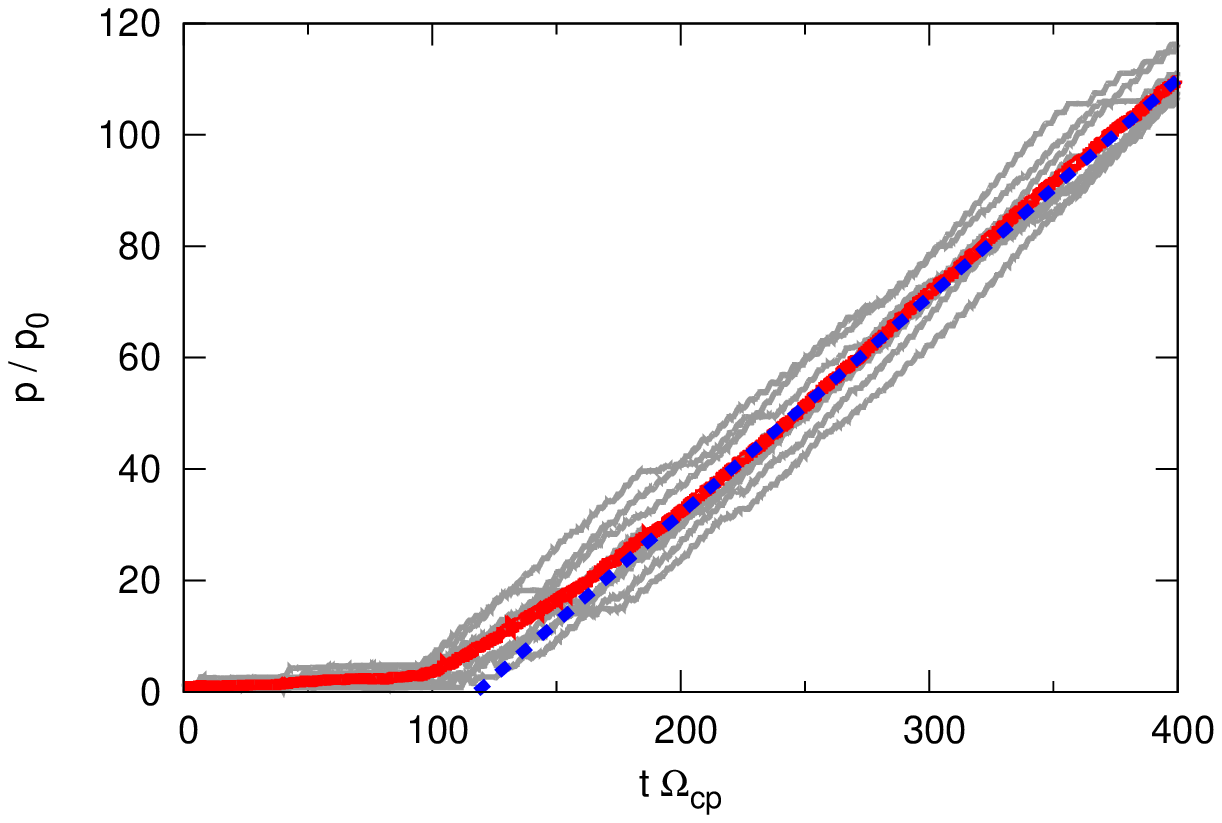}
\caption{Time evolution of momentum for ten most energetic particles (gray curves) and their average (red curve). The blue dashed line shows Equation~(\ref{eq:p}) with $\alpha=0.39$. 
\label{fig:6}}
\end{figure}

Figure~\ref{fig:4} shows the energy spectra of protons in the region of $2000~c/\omega_{\rm pp}\le x\le 7000~c/\omega_{\rm pp}$ at $t=400~{\Omega_{\rm cp}}^{-1}$. 
For the two-dimensional simulation (the black histogram), some protons are typically  accelerated to $10E_0$, where $E_0=m_{\rm p}v_{\rm d}^2/2$ is the initial kinetic energy. 
When leaking neutral particles are ionized in the upstream region, they are picked up by the upstream flow and accelerated to a few times $E_0$. 
Then, they are advected to the shock and accelerated by the second shock heating \citep{ohira13}. 
Even though there is strong magnetic turbulence, the diffusion perpendicular to the magnetic field line is strongly suppressed in the two-dimensional system. 
Therefore, the pickup ions are not injected to DSA in the two-dimensional simulation. 
The second shock heating can be interpreted as the SDA in the scatter-free system and accelerates particles only to about 4 times its energy. 
Hence, the typical energy of accelerated particles becomes $10~E_0$ for the two-dimensional simulation. 

On the other hand, for the three dimensional simulation (the red histograms), 
particles are very efficiently accelerated to $E\sim10^4~E_0~(v\sim10^2~v_{\rm d}\sim 0.3~c)$ at $t=400~\Omega_{\rm cp}^{-1}$ and the maximum energy is still increasing, where $v$ is the velocity. 
In the downstream region, the energy fractions of nonthermal particles with $E>4E_0$ and $E>10^3E_0$ are about $30\%$ and $1\%$ of the total kinetic energy, respectively. 

In order to understand the acceleration mechanism in the three-dimensional simulation, we show trajectories of an accelerated particle in Figure~\ref{fig:5}, where 
the panel (a) shows the time evolution of the kinetic energy and the panels (b)-(d) show the particle trajectories by assuming no periodic boundaries. 
The black curve in the panel (b) represents the mean position of the shock front.  
As one can see in the panel (b) of Figure~\ref{fig:5}, 
the particle interacts with the shock at $t\sim 10~\Omega_{\rm cp}^{-1}$, 
becomes a neutral hydrogen atom by the charge exchange process, 
and leaks into the upstream region. 
At $t\sim 60~\Omega_{\rm cp}^{-1}$, the leaking hydrogen atom is ionized and picked up by the upstream flow. 
The pickup ion interacts with the shock front again at $t\sim 100~\Omega_{\rm cp}^{-1}$. 
For the two-dimensional simulation, the particle is not accelerated further and just advected to the downstream region, while the particle is trapped around the shock front and accelerated for the three-dimensional simulation. 
After $t\sim 100~\Omega_{\rm cp}^{-1}$, the orbit in the upstream region is almost a simple gyro motion in the uniform magnetic field and the particle crosses the shock front at every gyro motion. 
The energy gain is correlated with the motion in the -z direction that is 
the same as the direction of the motional electric field, $\mbox{\boldmath $E$}_{\rm m}=-(\mbox{\boldmath $v$}_{\rm d}/c)\times\mbox{\boldmath $B$}_0$. 
Therefore, the acceleration mechanism is SDA. 
Contrary to the two-dimensional system, particles can diffuse in the direction perpendicular to the magnetic field line in the three-dimensional system. 
As a result, particles can go back to the upstream region and SDA with scattering can accelerate particles for a long time, 
that can be interpreted as DSA at the perpendicular shock.  
We track trajectories of 500 most energetic particles and confirm 
that all the energetic particles are accelerated by DSA at the perpendicular shock after they leak into the upstream region by the charge exchange process.
Hence, our simulation shows that the injection mechanism to DSA at the perpendicular shock in partially ionized plasmas is the leakage of hydrogen atoms by the charge exchange process. 

For SDA, during one gyro motion, $\Delta t\sim \Omega_{\rm cp}^{-1}$, the accelerated particles drift a distance of about $r_{\rm g}=v/\Omega_{\rm cp}$ in the direction of the motional electric field, so that the energy gain per gyro period is $\Delta E\sim |eE_{\rm m}|r_{\rm g}$. 
Then, the acceleration rate is given by
\begin{equation}
\frac{dE}{dt} \approx \frac{\Delta E}{\Delta t}=\alpha p_0 v\Omega_{\rm cp}~~,
\end{equation}
where $p_0=m_{\rm p}v_{\rm d}$ and $\alpha$ are the initial upstream momentum and a numerical factor of the order of unity, respectively. 
From the above equation, the momentum increasing rate is represented by 
\begin{equation}
\frac{dp}{dt} = \alpha p_{\rm 0}\Omega_{\rm cp}~~,
\label{eq:dpdt}
\end{equation}
and the solution is given by
\begin{equation}
p = \alpha p_0 \Omega_{\rm cp}(t-t_{\rm s}) + p_{\rm s}~~,
\label{eq:p}
\end{equation}
where $p_{\rm s}$ is a momentum at the start time of SDA, $t_{\rm s}$. 
Figure~\ref{fig:6} shows the time evolution of momentums for ten most energetic particles (gray curves) and their average (red curve). 
After the initial pickup process ($t > 100~\Omega_{\rm cp}^{-1}$), 
particles are accelerated and their momentum increases with time linearly 
as predicted by Equation~(\ref{eq:p}). 
Hence, the momentum evolution is consistent with SDA. 
$\alpha=0.39$ is derived by the fit of Equation~(\ref{eq:p}) to the average curve and the fitted line is shown by the blue dashed curve in the Figure~\ref{fig:6}. 

It should be noted that the ionization time scale is artificially lowered in this simulation. 
For actual young SNRs, the time scale of the initial pickup process becomes typically $10^5~\Omega_{\rm cp}^{-1}\approx 10^7~{\rm sec}$. 

\section{DISCUSSION}
\label{sec:4}

\begin{figure}
\includegraphics[scale=0.7]{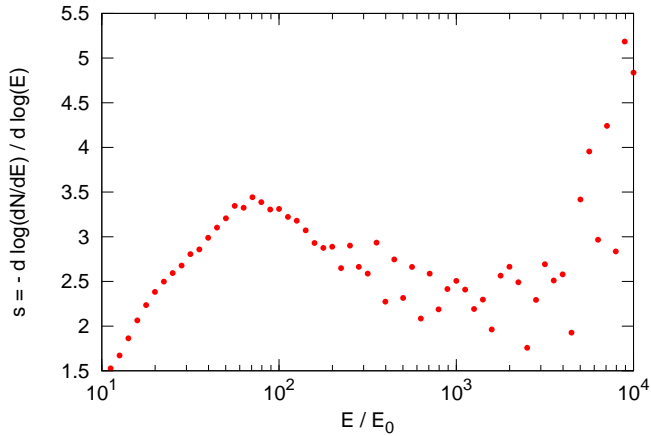}
\caption{Spectral index of the energy spectrum, $s=-d \log(dN/dE)/d \log E$, that is obtained from Figure~\ref{fig:4}.  
\label{fig:7}}
\end{figure}

In this section, we first discuss the acceleration time scale. 
Since ten most energetic particles in this simulation are still accelerating at $t= 400~\Omega_{\rm cp}^{-1}$, we can expect a further acceleration.  
From Equation~(\ref{eq:dpdt}), the acceleration time scale of SDA observed in this simulation is given by
\begin{eqnarray}
t_{\rm acc} &\equiv& \frac{p}{dp/dt} =\frac{1}{\alpha \Omega_{\rm cp}}\left(\frac{p}{p_0}\right) \nonumber \\
&=& 2.7~{\rm kyr}\left(\frac{\alpha}{0.4}\right)^{-1} \left(\frac{v_{\rm sh}}{10^3~{\rm km/s}}\right)^{-1} \nonumber \\
&& ~~~~~~~~~~\times \left(\frac{B}{3{\rm \mu G}}\right)^{-1} \left(\frac{E}{10^{15.5}{\rm eV}}\right)~~,
\label{eq:tacc}
\end{eqnarray}
where $v_{\rm d}\approx v_{\rm sh}$ and $p\approx E/c$ are used in the last equation. 
This is about 100 times smaller than that of DSA at the parallel shock. 
As already mentioned, SDA with scattering can be interpreted as DSA at the perpendicular shock. 
In fact, the minimum acceleration time scale of DSA at the perpendicular shock \citep {jokipii87} is the same as Equation~(\ref{eq:tacc}). 
Since the deceleration time scale of SNRs from the free expansion velocity ($\sim 10^4~{\rm km/s}$) to $10^3~{\rm km/s}$ is about $10~{\rm kyr}$ and the SNR size  at that time ($\sim 10~{\rm pc}$) is larger than the gyroradius of the knee ($r_{\rm g,knee}\approx 1~{\rm pc} (B/3{\rm \mu G})^{-1}$), CRs could be accelerated to the knee by SDA during the Sedov phase of SNRs. 
It should be noted that two conditions have to be fulfilled to extrapolate Equation~(\ref{eq:tacc}) to the knee. 
One is that the coherent length scale of the ordered magnetic  field in the ISM have to be larger than the gyro radius of the knee, $r_{\rm g,knee}$. 
The other is that particles with $E=10^{15.5}~{\rm eV}$ have to be efficiently scattered by magnetic turbulence in the downstream region in order to go back to the upstream region. 
Since our simulation is too small, 
our simulation has not shown yet that the later condition is fulfilled. 

Next, we discuss the spectral index of accelerated particles. 
The spectral index of accelerated particles, $s=-d \log(dN/dE)/d \log E$, in the nonrelativistic energy range is $s=1.5$ for the standard DSA at the parallel shock \citep{bell78,blandford78}. 
Figure~\ref{fig:7} shows the spectral index that is obtained from Figure~\ref{fig:4}. 
As shown in Figures~\ref{fig:4} and \ref{fig:7}, the energy spectrum is not a simple power law and the index is  $s=2-2.5$ around $E=10^3 E_0$. 
According to \citet{takamoto15}, the spectral index in the nonrelativistic energy range is $s\approx 2$ for the most efficient case of DSA at the test particle perpendicular shock ($\lambda_{\rm m.f.p}/r_{\rm g} =v/v_{\rm sh}$),  where $\lambda_{\rm m.f.p}$ is the mean free path of accelerated particles. 
Although the spectral index in this simulation seems to be consistent the test particle limit, the velocity structure around the shock front is strongly modified by the leaking neutral particles \citep{blasi12,ohira12,ohira13,ohira16} and 
the spectral index around $E=10^3 E_0$ is affected by the cutoff around $E\sim 10^4~E_0$ due to a finite time.  
In order to obtain the spectral index more precisely, we need a longer time simulation with a more realistic ionization time scale.

Finally, we discuss some influences by limitations of this simulation. 
In this simulation, we use artificially large cross sections of several ionization processes in order to reduce the ionization length scale and time scale. 
For actual SNR shocks, the length scale of the neutral precursor region is about $10^5~(v_{\rm s}/v_{\rm A}) c/\omega_{\rm pp}$ in the typical ISM, that is much larger than that of this simulation, $\sim10^3~c/\omega_{\rm pp}$. 
Since the gyroradius of protons with $E=10^4 E_0$ is also $r_{\rm g}\sim 10^3~c/\omega_{\rm pp}$ in the upstream region of this simulation, the spectrum of accelerated particles is affected by the artificial small precursor. 
Furthermore,  the smaller precursor in this simulation might result in a stronger turbulence artificially. 
In addition, the size of the simulation box transverse to the shock normal, $L_{\rm y}=L_{\rm z}=200~c/\omega_{\rm pp}$ is smaller than the gyroradius of accelerated protons in the upstream region. 
If we use a larger simulation box, the turbulence with larger length scale would be excited and the accelerated particles would be efficiently scattered in the upstream region. 
If so, the accelerated particle diffuse to the upstream region and the acceleration time scale becomes longer than that observed in this simulation. 
Furthermore, the injection to DSA and the acceleration time scale should be depend on many parameters (ionization fraction, electron temperature, and magnetic field orientation). 
Since perfect perpendicular shocks are never realized in astrophysical environments, it would be interesting to understand whether the results in this paper can be applied to oblique shocks or not. 
In order to studies above problems, we need to perform a more realistic large simulation and systematic studies. These will be addressed in future work. 

\section{SUMMARY}
\label{sec:5}
In this paper, we have performed the first three-dimensional hybrid simulation of  collisionless shocks generated by SNRs in partially ionized plasmas. 
The simulation results can be summarized as follows: 
\begin{itemize}
\item Like in our previous simulation, some downstream hydrogen atoms leak into the upstream region and modify the shock structures. 

\item Large magnetic field and density fluctuations are excited both in the upstream and downstream regions. They have three-dimensional structures. 

\item The leaking hydrogen atoms are accelerated by the pickup process in the upstream region. Then, they are injected to DSA (or SDA) at the perpendicular shock. 

\item Particles are accelerated to $v\sim100~v_{\rm sh}\sim0.3~c$ at $t=400~\Omega_{\rm cp}^{-1}$. The acceleration rate in this simulation is faster than 
that of the typical DSA, up to the end of the simulation. 
\end{itemize} 
Isolated SNRs (e.g. SNRs of type Ia supernovae) are expected to be generated frequently in the partially ionized ISM. 
Therefore, our results show that such SNRs can produce GCRs and suggest that such SNRs can accelerate CRs to the knee energy if the shock is a perpendicular shock. 

\acknowledgments
Numerical computations were carried out on the XC30
system at the Center for Computational Astrophysics (CfCA)
of the National Astronomical Observatory of Japan.
We thank T. Terasawa, T. Inoue and R. Yamazaki for useful comments. 
We are also grateful to the anonymous referee for useful suggestions and comments. 
This work was supported in part by Grants-in-Aid for Scientific Research of the Japanese Ministry of Education, Culture, Sports, Science and Technology No. 16K17702.

\end{document}